\documentclass[aps,pra,twocolumn,showpacs]{revtex4}
\usepackage{color, graphicx}
\usepackage{amsmath, amssymb}

\definecolor{rosa}{cmyk}{0,1,0.50,0}

\begin{document}

\title{Quasiprobabilities for Multipartite Quantum Correlations of Light}

\author{E. Agudelo} \email{elizabeth.ospina@uni-rostock.de}\affiliation{Arbeitsgruppe Theoretische Quantenoptik, Institut f\"ur Physik, Universit\"at Rostock, D-18051 Rostock, Germany}
\author{J. Sperling}\affiliation{Arbeitsgruppe Theoretische Quantenoptik, Institut f\"ur Physik, Universit\"at Rostock, D-18051 Rostock, Germany}
\author{W. Vogel}\affiliation{Arbeitsgruppe Theoretische Quantenoptik, Institut f\"ur Physik, Universit\"at Rostock, D-18051 Rostock, Germany}
\pacs{42.50.Ct, 03.65.Ta,  03.65.Wj, 03.67.Bg}

\begin{abstract}
	Regular quasiprobabilities are introduced for the aim of characterizing quantum correlations of multimode radiation fields.
	Negativities of these quantum-correlation quasiprobabilities are necessary and sufficient for any quantum-correlation encoded in the multimode Glauber-Sudarshan $P$~function.
	The strength of the method is demonstrated for a two-mode phase randomized squeezed-vacuum state. It has no entanglement, no quantum discord, a positive Wigner function, and a classical reduced single-mode representation. Our method clearly visualizes the quantum correlations of this state.
\end{abstract}
\date{\today}
\maketitle

\section{Introduction}
\label{Sec:Introduction}
	Currently a variety of different notions of quantumness are discussed.
        It is of fundamental interest to classify the nature into one part showing classical signatures, and another one dominated by quantum phenomena.
	In quantum optics, the most common notion of nonclassicality is esta\-blished through the Glauber-Sudarshan $P$~function~\cite{glauber63,sudarshan63}.
	If this distribution does not resemble a classical probability density, the corresponding state is called a nonclassical one~\cite{titulaer65,mandel86}.

	Quantum entanglement is another notion of quantumness~\cite{horodecki09, guehne09}.
	Although each indivi\-dual subsystem can be quantum, for a separable state the correlations between the subsystems can be des\-cribed by classical statistics~\cite{werner89}.
	Entanglement or inseparability can be fully characterized by negativities of optimized quasiprobability (QP) distributions~\cite{sperling09, sperling12}, denoted as $P_{\rm Ent}$.

	In quantum information theory, a quantum-correlation (QC) is also characterized by the so-called quantum discord~\cite{ollivier01, henderson01}.
	A nonzero discord describes the back action of a measurement in one subsystem to another one.
	Discord is often considered as a general measure of QC beyond entanglement. 
	Recently, however, it has been shown that the notions of QC based on quantum discord and on negativities of the $P$~function are maximally inequivalent~\cite{ferrarro12}.

	The general characterization of the QCs of radiation fields has been formulated in terms of the space-time dependent $P$~functional~\cite{vogel08}.
	This yields a full hierarchy of inequalities for observable correlation functions.
	Whenever the described systems violate classical probability theory, the $P$~functional is a QP representation of quantum light.
	In the following we restrict attention to equal-time measurements, so that the $P$~functional simplifies to the $n$-partite $P$~function of the global quantum state $\hat\rho$,
	\begin{align}
		\hat\rho =\int {\rm d}^{2n}\boldsymbol\alpha\, P(\boldsymbol\alpha)\,|\boldsymbol\alpha\rangle\langle \boldsymbol\alpha|,
	\end{align}
	where $|\boldsymbol\alpha\rangle=|\alpha_1,\dots,\alpha_n\rangle$ denotes multimode products of coherent states.
	One severe disadvantage of the $P$~function consists in its strong singularities occurring for many quantum states, even in the single-mode case.
	As a consequence, in general this function is experimentally not accessible and hence of limited practical value.
	Only in special cases the $P$~function is regular~\cite{agarwal92} and can be measured~\cite{kiesel08}.

	The concept of phase space representations has been generalized to the $s$-parametrized QPs~\cite{cahill69}.  
	The latter include popular examples: the $P$~function ($s=1$), the Wigner function ($s=0$), and the Husimi~$Q$~function ($s=-1$).
	Diminishing the  $s$~parameter, the QPs become more regular, but less sensitive for testing quantumness.
	For squeezed states, all $s$-parametrized QPs are either positive or irregular.
	Especially the Wigner function is popular, since it is easily obtained in experiments, e.g. for quantum light, molecules, and trapped atoms~\cite{smithey93,dunn95, ourjoumtsev06, Deleglise08,leibfried96}.
	Further generalizations of the QP methods were introduced in~\cite{agarwal70}.

	For the single-mode case, nonclassicality QPs, $P_{\rm Ncl}$, have been introduced~\cite{kiesel10}.
	They are regularized versions of the highly singular $P$~functions.
	For any nonclassical single-mode state, they show negativities and can be directly obtained from experimental data~\cite{kiesel11,kiesel11a,kiesel12}.
	Hence $P_{\rm Ncl}$ is a powerful tool for the full experimental characterization of quantum effects of single-mode fields.

	In the present paper, we generalize the regularization of the $P$ function for multimode light.
	The resulting QP distribution, $P_{\rm QC}$, uncovers any QC occurring in the multimode $P$~function. 
	For practical applications it is  of great importance that  $P_{\rm QC}$ can be directly sampled from experimental data obtained by multimode homodyne detection. 
	We show that our method uncovers QCs contained in a family of highly singular $P$~functions, which exhibit QC's beyond quantum discord and quantum entanglement.

	The paper is structured as follows. 
	We provide a scenario for use of our method in Sec.~\ref{Sec:Motivation} through the full phase-randomized two-mode squeezed vacuum state, which is not entangled, has zero quantum discord, a positive Wigner function, and classical reduced subsystems.
	In Sec.~\ref{Sec:MultipartitePQCfunction} we present the regularization of the $P$ function for multimode radiation fields.
	The resulting QP distribution, $P_{\rm QC}$, uncovers any QC occurring in the multimode $P$~function.
	We show in Sec.~\ref{Sec:BeyondEntanglementDiscord} that for this special state, our method uncovers QCs contained in the highly singular $P$~function, via negativities of the bipartite filtered QP distribution $P_{\rm QC}$. 
	The QCs are clearly visible, even when the other signatures of QCs do not persist.
	In Sec.~\ref{Sec:Sampling} we provide the approach of direct sampling of $P_{\rm QC}$ from experimental data obtained by multimode homodyne detection.
	A summary and some conclusions are given in Sec.~\ref{Sec:Conclusions}.

\section{Motivation}
\label{Sec:Motivation}
\begin{figure}[ht]
	\centering
	\includegraphics*[width=5cm]{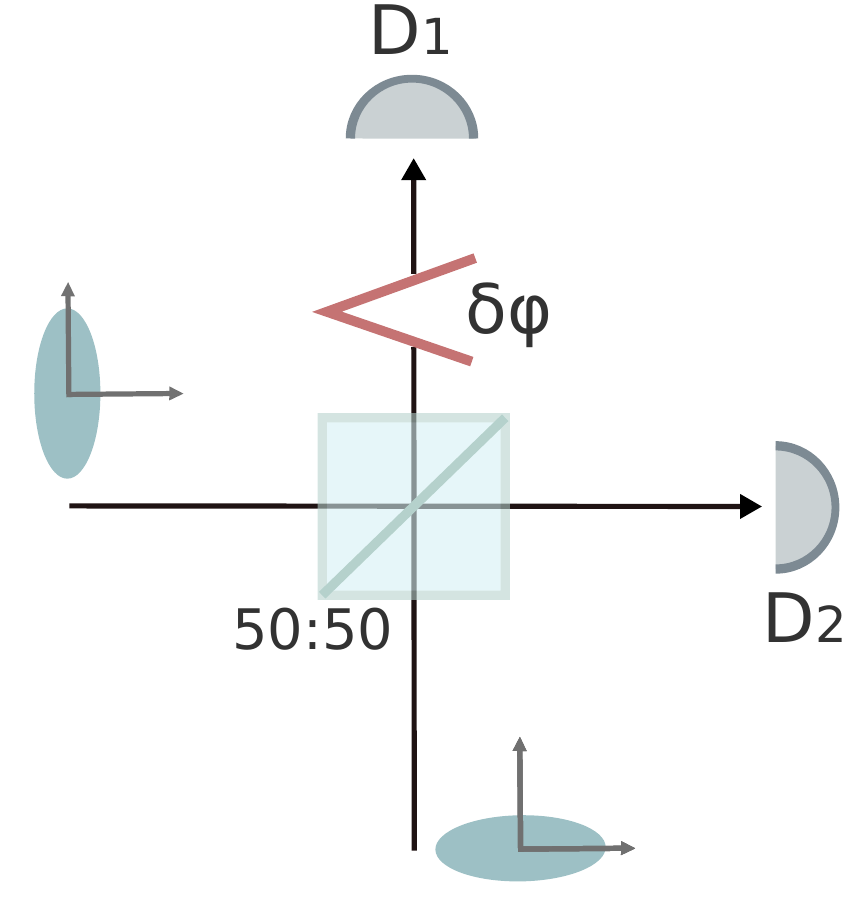}
	\caption{(Color online) Experimental setup for the generation of a phase randomized two-mode squeezed vacuum.
		A 50:50 beam splitter combines squeezed-vacuum states,  which are squeezed in orthogonal quadratures.
		One of the output channels is fully phase randomized, $\delta\varphi=2\pi$.}
	\label{experiment}
	\end{figure}

	Let us consider a realistic experimental scenario for generating a bipartite continuous variable state.
	The inputs of a 50:50 beam splitter are equally squeezed states in orthogonal quadratures, cf. Fig. \ref{experiment}. 
	In the output ports of this setup one obtains a two-mode squeezed vacuum (TMSV) state.
	Such states are entangled and have non-zero quantum discord.

	The situation is drastically changed, if phase randomization, indicated by $\delta\varphi$ in Fig.~\ref{experiment}, occurs in one of the output ports.
	The entanglement properties of the state depend sensitively on the dephasing~\cite{sperling12}. In the case of an equally distributed phase, $\delta\varphi=2\pi$, the resulting output state is a fully phase randomized TMSV state,
	\begin{align}
		\hat\rho=\sum_{n=0}^\infty (1-p)p^n |n\rangle\langle n|\otimes|n\rangle\langle n|,
		\label{Eq State}
	\end{align}
	with $0<p<1$, $p$ being related to the squeezing parameter of the initial input fields.

	It is important to consider properties of this state which are closely related to its QCs.
	First, the phase randomized TMSV state is a convex mixture of the product states $|n\rangle\langle n|\otimes|n\rangle\langle n|$.
	Hence, it is classical with respect to the property of entanglement. 
	Second, due to the orthogonality of the Fock states, $\langle n|n'\rangle=0$ for $n\neq n'$, this state has no QCs in the sense of the quantum discord. 
	This conclusion is obvious from the general representation of quantum states with zero discord given in~\cite{datta10}.
	So far our state belongs to the class of quantum states considered in~\cite{ferrarro12} with the aim to demonstrate the inequivalence of QCs based on the $P$~function and the quantum discord. 
	In our context, we require the following additional properties.

\begin{figure}[ht]
	\centering
	\includegraphics*[scale=0.5]{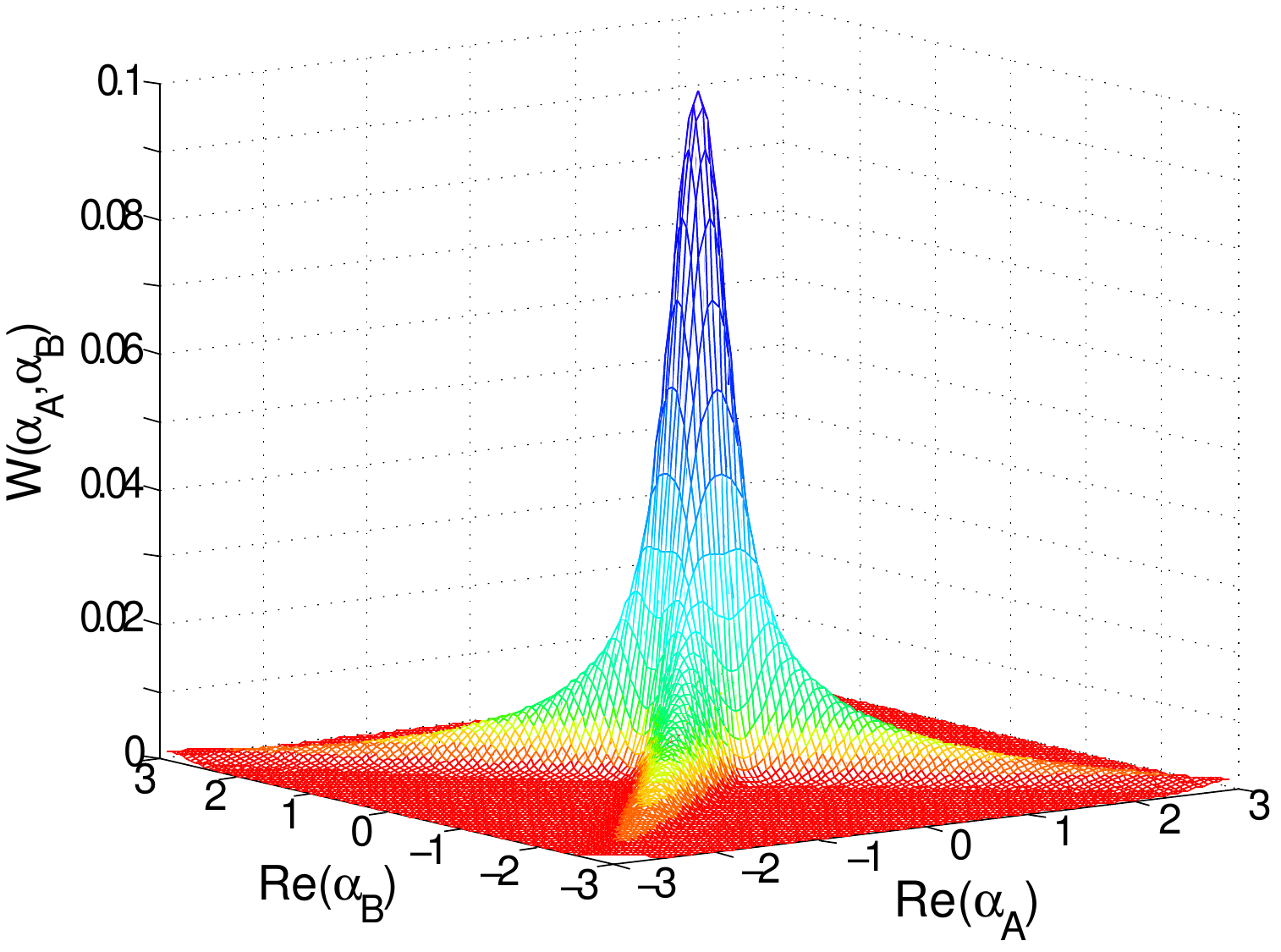}
	\caption{(Color online) The Wigner function is shown for a fully phase randomized two-mode squeezed vacuum state, with $p=0.8$.
	It is non-negative in the full two-mode phase space.}
	\label{wigner}
	\end{figure}

	Third, let us consider the reduced density operator, $\hat\rho_{\rm red}$. 
	Due to the symmetry of this state with respect to the interchange of both subsystems, we find that
	\begin{align}
		\hat\rho_{\rm red}={\rm Tr}_A\hat\rho={\rm Tr}_B\hat\rho=\sum_{n=0}^\infty (1-p)p^n |n\rangle\langle n|.
	\end{align}
	This state is a thermal one with a mean photon number $\bar n_{\rm th}=p/(1-p)$. 
	Therefore it shows a classical behavior with respect to the reduced single-mode states.
	This property insures that any identified signature of quantumness exposes a true QC effect. 
	Fourth, the full state $\hat\rho$ is a mixture -- due to the phase randomization -- of two-mode Gaussian states.
	Hence it has a non-negative Wigner function, which is shown in Fig.~\ref{wigner}. 
	This QP distribution, which was often applied in experiments~\cite{smithey93,dunn95, ourjoumtsev06, Deleglise08,leibfried96}, is improper to visualize QCs of this state
        by attaining negative values.

	Summarizing the features of the fully phase randomized TMSV state, we have introduced an experimentally realizable quantum state with the following properties:
	\begin{enumerate}
        \item no entanglement;
	\item zero quantum discord;
	\item classical reduced single-mode states;
	\item non-negative Wigner function.
	\end{enumerate}
        Despite these strong signatures of classicality, this state will be proven to show clear QC effects.

	It is a non-Gaussian state that can be experimentally generated as outlined above. 
	To prove that this state describes QCs, we have to visualize the negativities of its strongly singular, two-mode $P$~function:
        \begin{align}
		\nonumber P(\alpha_A,\alpha_B)&=\sum_{n=0}^\infty (1-p) p^n \sum_{k,l=0}^n \binom{n}{k}\binom{n}{l}\, (-1)^{k+l} \\
		&\times \frac{n!}{k!\,l!}\, \partial^k_{\alpha_A}\partial^k_{\alpha_A^*} \delta(\alpha_A)\, \partial^l_{\alpha_B}\partial^l_{\alpha_B^*} \delta(\alpha_B).
		\label{eq:P-rand}
	\end{align}
	We will shown that there are QCs between the subsystems $A$ and $B$ going beyond entanglement and quantum discord.

\section{Regularized multipartite $P_{\rm QC}$~function}
\label{Sec:MultipartitePQCfunction}
	To be useful for experiments, quantumness criteria must be based on well-behaved functions.
	Hence a regularization of the highly singular Glauber-Sudarshan $P$~representation is required.
	We can construct a multimode regular $P_{\rm QC}$ function in the form of a convolution
	\begin{align}\label{Eq:PQC-Def}
		P_{\rm QC}(\boldsymbol\alpha; w)=\int {\rm d}^{2n}\boldsymbol\alpha'\, P(\boldsymbol\alpha-\boldsymbol\alpha')\,\tilde\Omega_w(\boldsymbol\alpha'),
	\end{align}
	where $\tilde\Omega_w$ is a suitable function, which we are going to construct.
	The occurring width parameter $w$ provides the property $P_{\rm QC}(\boldsymbol\alpha; w)\to P(\boldsymbol\alpha)$ for $w\to\infty$. A direct sampling formula for measured quadrature data will be given.

	We start from the multimode characteristic function.
	The $n$-mode characteristic function $\Phi$ is defined as the Fourier transform of the $P$~function
	\begin{align}
		\Phi(\boldsymbol\beta) =\int {\rm d}^{2n}\boldsymbol\alpha\, P(\boldsymbol\alpha)\, {\rm e}^{\boldsymbol\beta\cdot\boldsymbol\alpha^*-\boldsymbol\beta^*\cdot\boldsymbol\alpha}.
	\end{align}
	The characteristic function $\Phi$ is always a continuous function, independent of singularities in $P(\boldsymbol\alpha)$. 
	Single-mode nonclassicality criteria based on the characteristic function have been introduced~\cite{vogel00,Ri-Vo02} and applied~\cite{CF-exp1,CF-exp2,kiesel09}.
	Note that the $n$-mode characteristic function is bounded in the form $|\Phi(\boldsymbol\beta)|\leq \exp(|\boldsymbol\beta|^2/2)$.

	The convolution in Eq.~\eqref{Eq:PQC-Def} is a point-wise product in Fourier space.
	Hence, we consider the so-called multimode filtered characteristic function $\Phi_{\rm QC}(\boldsymbol\beta;w)$,
	\begin{align}
		\Phi_{\rm QC}(\boldsymbol\beta;w)=\Phi(\boldsymbol\beta) \prod_{k=1}^n\Omega(\beta_k/w),
	\end{align}
	being the Fourier transformed of $P_{\rm QC}$ for $0<w<\infty$.
	Our choice for the filter $\Omega(\beta)$ is the autocorrelation function
	\begin{align}\label{Eq:SingleModeFilter}
		\Omega(\beta)=\left(\frac{2}{\pi}\right)^{3/2}\int {\rm d}^2\beta'\,{\rm e}^{-|\beta+\beta'|^4}{\rm e}^{-|\beta'|^4}.
	\end{align}
	Finally, we get the regularization kernel in Eq.~\eqref{Eq:PQC-Def} by the inverse Fourier transform as
	\begin{align}\label{Eq:ConvolutionKernel}
		\tilde\Omega_w(\boldsymbol\alpha)= \prod_{k=1}^n \left(\frac{1}{\pi^2}\int {\rm d}^2\beta\,\,\Omega(\beta/w)\, {\rm e}^{\beta^\ast\alpha_k-\beta\alpha_k^\ast} \right).
	\end{align}

	For the single-mode case, such a filter $\Omega (\beta/w) $ in Eq.~\eqref{Eq:SingleModeFilter} has been introduced and characterized in Ref.~\cite{kiesel10}.
	It has been shown that this filter suppresses the exponentially rising behavior of $\exp(|\beta|^2/2)$.
	In addition, it has been shown that this kind of filter belongs to the class of invertible filters, $\Omega(\beta/w)\neq 0$.
	Hence, deconvolution of Eq.~\eqref{Eq:PQC-Def} yields the $P$~function, and thus the full state $\hat\rho$, from the QP~distribution $P_{\rm QC}$.

	Let us comment here on the structure of the regularizing function $\tilde\Omega_w$ defined in Eq.~(\ref{Eq:ConvolutionKernel}).
	This multimode function can be written as a product of single-mode functions which do not depend on the state, the proof is given in Appendix~A.
	A product filter is a practical tool that is sufficient to identify any kind of nonclassicality in the multimode $P$~function.
	The latter allows us directly to recognize QCs in quantum systems with classical parts.

	The regularizing function must not introduce any quantum-correlation to $P_{\rm QC}$, which is absent in the Glauber-Sudarshan $P$~function.
	For any nonclassical state which includes some QC, there exist values $w$ and $\boldsymbol\alpha$ for which $P_{\rm QC}(\boldsymbol\alpha; w)<0$.
	The other way around, a state is classical, if for all values $w$ and $\boldsymbol\alpha$ the function $P_{\rm QC}(\boldsymbol\alpha; w)$ represents a classical probability density.
	This means that we may identify QCs via uncorrelated filtering.

\section{Beyond entanglement and discord}
\label{Sec:BeyondEntanglementDiscord}

\subsection{Direct verification of quantum correlations}
\begin{figure}[ht]
	\centering
	\includegraphics[scale=0.45]{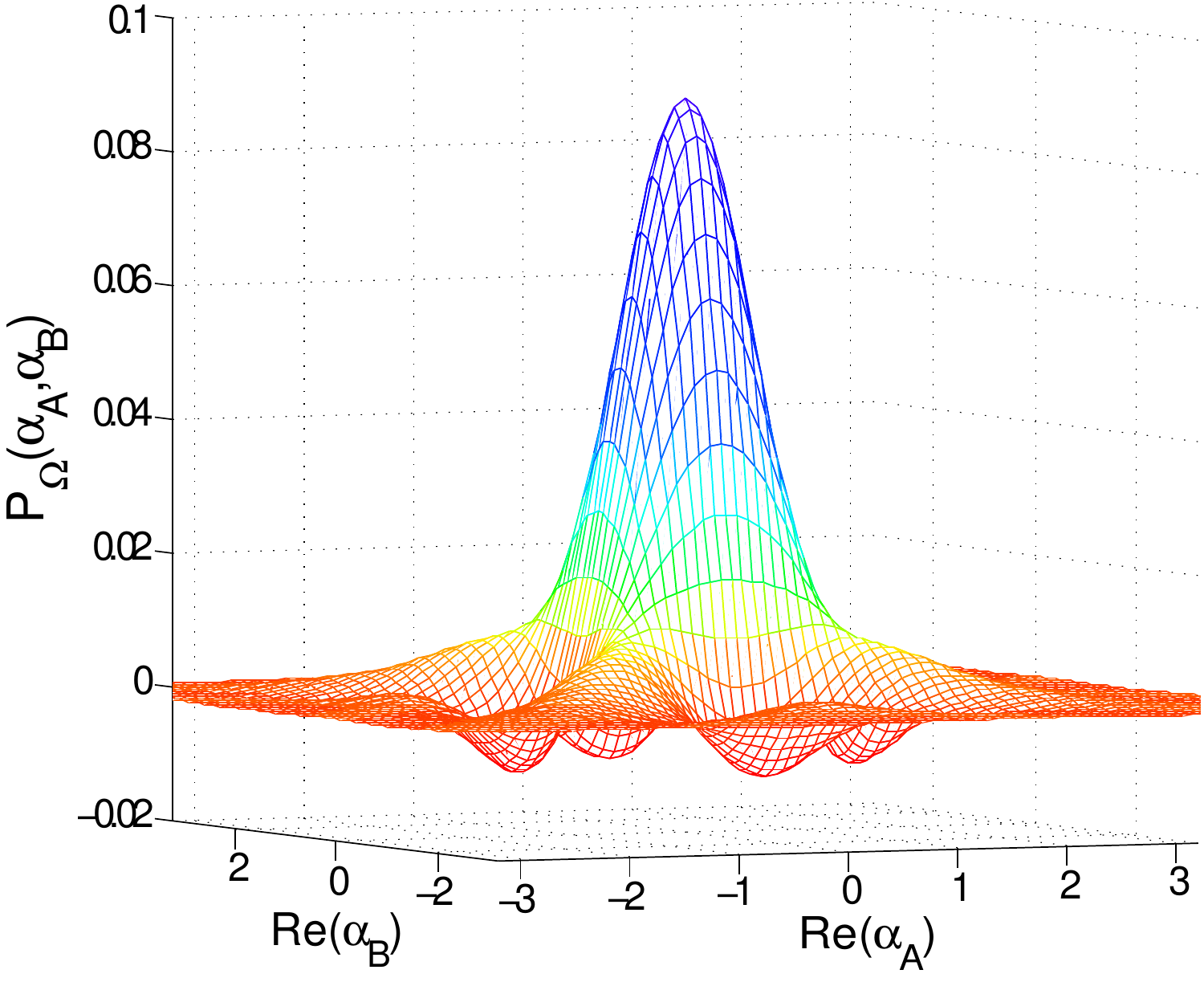}
	\caption{(Color online) The regularized quasiprobability $P_{\rm QC}$ is shown for a fully phase randomized two-mode squeezed vacuum state, for $p=0.8$ and $w=1.5$.
		Clear negativities visualize the nonclassical correlations of the state.}
	\label{pfil}
	\end{figure}

        For the proof of the existence of QCs in the phase randomized TMSV state, the regularized two-mode quantum-correlation QP, $P_{\rm QC}$, is calculated from Eqs. ~(\ref{eq:P-rand}), (\ref{Eq:PQC-Def}), (\ref{Eq:SingleModeFilter}), and (\ref{Eq:ConvolutionKernel}).
	The result in Fig.~\ref{pfil} clearly shows that $P_{\rm QC}$ becomes  negative for properly chosen arguments.
	In view of the properties of the state listed above, these negativities are a direct proof of the quantum nature of the correlations between the two subsystems $A$ and $B$. Although the 
	Wigner function in Fig.~\ref{wigner} contains the full information on the quantum state, it does not directly visualize the QC properties of the state under study.
	The regular QP distribution $P_{\rm QC}$, on the other hand, directly displays the QCs within the considered randomized TMSV state by attaining negative values.

	Here the question may arise of whether the quantum correlations of the fully phase randomized TMSV state can be used in any application.
	The heralded control of the number of photons in an arbitrarily chosen time interval can be done with these kinds of quantum-correlated states. In our case, for sufficiently strong squeezing the control could be extended to higher photon numbers compared with the often used heralding via spontaneous parametric down conversion.
        More generally, even the time sequence of a train of photons could be controlled to some extend by the quantum correlations inherent in such a state.

\subsection{Identifying general quantum correlations}
	Now we consider the characterization of QC for arbitrary states in a more general context.
        Negative values of $P_{\rm QC}$ also include negativities which can be explained by entanglement, quantum discord, or single-mode nonclassicalities.
	Let us outline how one can identify those particular effects in general.
	
	A nonzero quantum discord may be directly or numerically computed~\cite{giorda}.
	For entanglement, one can get the entanglement QP $P_{\rm Ent}$, e.g. for the considered phase-randomized TMSV state it is given in Ref.~\cite{sperling12}.
	It is worth mentioning that a nonclassical multimode $P$~function is necessary to describe an entangled state~\cite{sperling09}.
	As our example shows, however, a negative $P_{\rm QC}$ is not sufficient to verify entanglement.

	In a last step, one can try to find negativities of $P_{\rm QC}$ which may be given by single-mode nonclassicalities.
	Here one has to consider the reduced states.
	These reduced states can be characterized by the single-mode nonclassicality QP, $P_{\rm Ncl}$, introduced in~\cite{kiesel10}.
	More generally, one may consider $n$-mode states which are fully characterized by $P_{\rm QC}(\alpha_1,\dots,\alpha_n;w)$.
	Tracing over some subsystems, we get a state which is described by the remaining subsystems only, e.g.:
	\begin{align}
		P_{\rm QC}(\alpha_1,\dots,\alpha_{n-1};w)=\int {\rm d}^2\alpha_n\,P_{\rm QC}(\alpha_1,\dots,\alpha_n;w).
	\end{align}
	If the original state is quantum correlated but none of the partially traced states is quantum correlated, one may refer to this kind of correlation as genuine $n-$mode~QC.

	The multimode analysis does not depend on the kind of mode decomposition of the radiation field.
	We may consider a two-mode scenario, a similar processing can be done in the multimode case.
	The individual modes may be represented by the corresponding annihilation operators $\hat a_{A}$ and $\hat a_{B}$.
	In addition, the system may be decomposed in two other modes, $\hat a_{A'}$ and $\hat a_{B'}$.
	A mode transformation is given by a unitary transform $\boldsymbol U$ of the form
	\begin{align}
		\begin{pmatrix}
			\hat a_{A'}\\ \hat a_{B'}
		\end{pmatrix}
		=
		\begin{pmatrix}
			U_{A'A} & U_{A'B} \\
			U_{B'A} & U_{B'B}
		\end{pmatrix}
		\begin{pmatrix}
			\hat a_{A}\\ \hat a_{B}
		\end{pmatrix}.
	\end{align}
	One can easily see that such a transformation maps a coherent state $|\alpha_{A},\alpha_{B}\rangle$ to
	\begin{align}
		|\alpha_{A'},\alpha_{B'}\rangle=|U_{A'A}\alpha_{A}+U_{A'B}\alpha_{B},U_{B'A}\alpha_{A}+U_{B'B}\alpha_{B}\rangle,
	\end{align}
	which is also a coherent product state.
	Hence, we conclude that $P_{\rm QC}(\alpha_{A},\alpha_{B};w)$ has negativities for some $w$, iff $P_{\rm QC}(\alpha_{A'},\alpha_{B'};w')$ has negativities for some $w'$.
	Note that such an equivalence is, in general, not true for the reduced state in $A$ and $A'$ or $B$ and $B'$.

\section{Sampling of $P_{\rm QC}$}
\label{Sec:Sampling}
	For applications of our method, a direct sampling formula is useful which yields $P_{\rm QC}$ from the measured data.
	We consider multimode homodyne detection, which gives a sample of $N$ measured multimode quadrature values $\{( x_k[j], \varphi_k[j])_{k=1}^n\}_{j=1}^N$.
	The index $k$ denotes the mode and $j$ numbers the measured values for the quadrature $x_k[j]$ and its corresponding phase $\varphi_k[j]$.
	Since our multimode regularizing function $\tilde\Omega_w$ can be written as a product, we obtain the sampling formula
	\begin{align}\label{Eq:Sampling}
		P_{\rm QC}(\boldsymbol\alpha;w)=\frac{1}{N}\sum_{j=1}^N \prod_{k=1}^n f(x_k[j],\varphi_k[j],\alpha_k;w).
	\end{align}
	The needed single-mode pattern function reduces for phase randomized states to
	\begin{align}\label{Eq:patternfct}
		f(x,\alpha;w)=&\frac{1}{\pi^2}\int_{-\infty}^{+\infty} {\rm d}b\, b\, {\rm e}^{b^2/2}\, \Omega\left(b/w\right)\times\\
		&\int_{0}^{2\pi} {\rm d}\varphi\, {\rm e}^{ibx(\pi/2-\varphi)+2 i |\alpha| b \sin\left[\arg \alpha -\varphi\right]},\nonumber
	\end{align}
	and it has to be calculated only once.

	In order to provide the confidence of the negativity of the sampling in Eq.~\eqref{Eq:Sampling}, we need to estimate the statistical uncertainties.
	The empirical variance of the sampling, $\sigma^2\left[P_{\rm QC}(\boldsymbol\alpha;w)\right]$, is derived in the Appendix~B.
	The statistical uncertainty of the sampled estimate in Eq.~\eqref{Eq:Sampling} is
	\begin{align}
		\Delta[P_{\rm QC}(\boldsymbol\alpha;w)]=\frac{ \sigma\left[P_{\rm QC}(\boldsymbol\alpha;w)\right]}{\sqrt N}.
	\end{align}
	Thus one can obtain the confidence of a negativity as
	\begin{align}
		\mathcal{C}(\boldsymbol\alpha;w)= \frac{|P_{\rm QC}(\boldsymbol\alpha;w)|}{\Delta[P_{\rm QC}(\boldsymbol\alpha;w)]},
	\end{align}
	for a given $\boldsymbol\alpha$ with $P_{\rm QC}(\boldsymbol\alpha;w)<0$.
	One may choose the value $w>0$, such that $\mathcal{C}(\boldsymbol\alpha;w)$ becomes maximal.

\section{Summary and Conclusions}
\label{Sec:Conclusions}
	A regularized multimode version of the Glauber-Sudarshan $P$~function is introduced: the quantum-correlation quasiprobability $P_{\rm QC}$, which is parameterized only by a single width parameter $w$. 
        Through its negativities, it directly visualizes quantum correlations included in any multimode quantum state. 
        The negativities occur, if and only if the state has a nonclassical multimode Glauber-Sudarshan $P$~function. 
        The method of regularization is universal, since it does not depend on the considered state. 
	
	Our method is applied to a fully phase-randomized two-mode squeezed-vacuum state. 
	This state is shown to be classical with respect to other established notions of quantum correlations, it is not entangled and it has zero quantum discord. 
	The two-mode Wigner function is shown to be positive semidefinite. 
	Last but not least, the state is locally classical in each mode. However, the negativities in the bipartite quantum-correlation quasiprobability clearly uncover the quantum-correlation properties of this state, despite the disappearance of quantum entanglement and quantum discord.

        For the efficient application of our method in experiments, we explicitly provide a direct sampling formula. It is based on the data which are recorded in two-mode balanced homodyne detection.
        One may directly determine the quantum-correlation quasiprobabilities
        together with their statistical errors. On this basis it is straightforward 
        to estimate the statistical significance of the observed quantum-correlation effects.
        
\section*{Acknowledgments}
	The authors are grateful to T. Kiesel for enlightening discussions.
	We are also grateful to the anonymous referee for helpful comments.
	This work was supported by the Deutsche Forschungsgemeinschaft through SFB 652.

\appendix
\begin{widetext}
\section{Regularization}
	Here, we prove the properties of the multimode filtering procedure.
	For the single-mode scenario, we refer to the appendix in Ref.~\cite{kiesel10}.
	First, we show that the a product filter is sufficient to identify any kind of nonclassicality contained in the multimode $P$ function.
	Second, we show that the filtered quasiprobability $P_{\rm QC}$ is a smooth function.

\subsection{Product filters}
	For convenience, we will argue in the Fourier domain.
	The multimode filtered characteristic function (CF) reads as
	\begin{align}
		\Phi_{\rm QC}(\boldsymbol\beta; w)=\Phi(\boldsymbol\beta)\prod_{m=1}^n \Omega(\beta_m/w),
	\end{align}
	with $\Omega$ being the continuous function
	\begin{align}\label{Eq:SingleModeFilter2}
		\Omega(\beta)=\left(\frac{2}{\pi}\right)^{3/2}\int {\rm d}^2\beta'\,{\rm e}^{-|\beta+\beta'|^4}{\rm e}^{-|\beta'|^4},
	\end{align}
	with $\lim_{w\to\infty} \Omega(\beta/w)=1$.

	Bochner's theorem can be given in the following form~\cite{bochner33, vogel00}.
	The CF $\Phi_{\rm QC}$ is the Fourier transform of a probability distribution, iff for all finite sequences $f_k\in\mathbb C$ and all finite sequences $\boldsymbol\beta_k\in\mathbb C^n$ it holds that
	\begin{align}\label{Eq:PosCF}
		\sum_{k,k'} f_k^\ast f_{k'} \Phi_{\rm QC}(\boldsymbol\beta_k-\boldsymbol\beta_{k'}; w)\geq 0,
	\end{align}
	together with the normalization $\Phi_{\rm QC}(\boldsymbol 0; w)=1$ and the symmetry $\Phi_{\rm QC}(\boldsymbol \beta; w)^\ast=\Phi_{\rm QC}(-\boldsymbol \beta; w)$.	
	We may rewrite the matrix $(\Phi_{\rm QC}(\boldsymbol\beta_k-\boldsymbol\beta_{k'}; w))_{k,k'}$ in the following form:
	\begin{align}
		(\Phi_{\rm QC}(\boldsymbol\beta_k-\boldsymbol\beta_{k'}; w))_{k,k'}=(\Phi(\boldsymbol\beta_k-\boldsymbol\beta_{k'}))_{k,k'}\circ \Omega([\beta_{1,k}-\beta_{1,k'}]/w)_{k,k'}\circ\dots\circ\Omega([\beta_{n,k}-\beta_{n,k'}]/w)_{k,k'},
	\end{align}
	where $\circ$ denotes the entrywise or Schur product.
	It is well known that the Schur product of positive semidefinite matrices is again positive semidefinite, and the matrix $(1)_{k,k'}$ is the identity element of the Schur product.

	Let us assume that our state is a classically correlated one.
	This means that the multimode $P$~function is a non-negative probability density, which implies that $\Phi$ fulfills Bochner's theorem, cf. Eq.~\eqref{Eq:PosCF}, or simply $(\Phi(\boldsymbol\beta_k-\boldsymbol\beta_{k'}))_{k,k'}$ is a positive semidefinite matrix.
	From the definition of $\Omega$, see Eq.~(\ref{Eq:SingleModeFilter2}), it is clear that $\Omega([\beta_{m,k}-\beta_{m,k'}]/w)_{k,k'}$ (for the vector components $m=1,\dots,n$) is also positive semidefinite.
	In the case $w\to\infty$, we have $\Omega([\beta_{m,k}-\beta_{m,k'}]/w)_{k,k'}\to (1)_{k,k'}$.
	This means that for any $w>0$, we have a positive semidefinite matrix $(\Phi_{\rm QC}(\boldsymbol\beta_k-\boldsymbol\beta_{k'}; w))_{k,k'}$.

	Let us now consider a quantum-correlated state.
	Thus, the condition in Eq.~\eqref{Eq:PosCF} cannot be true.
	This means that there exists a finite sequence $f_k\in\mathbb C$ and a finite sequence $\boldsymbol\beta_k\in\mathbb C^n$, such that the quadratic form $(\Phi_{\rm QC}(\boldsymbol\beta_k-\boldsymbol\beta_{k'}; w))_{k,k'}$ is not positive semidefinite.
 	The continuity of $\Omega$ for $w\to\infty$ implies $\Omega([\beta_{m,k}-\beta_{m,k'}]/w)_{k,k'}\to (1)_{k,k'}$, we get that there exists a finite $w>0$ such that  $(\Phi_{\rm QC}(\boldsymbol\beta_k-\boldsymbol\beta_{k'}; w))_{k,k'}$ is not a positive semidefinite matrix.
	Thus, the inverse Fourier transformed $P_{\rm QC}(\boldsymbol\alpha; w)$ cannot be classical.
	This holds true for an interval of width parameters from the particular $w$ to $\infty$.

	Concluding this first part, we have seen that the product filter for a particular choice of $w$ exhibits any kind of quantum correlations within the state.
	The other way around, a state is classical, iff for any $w>0$, we have a non-negative $P_{\rm QC}$.

\subsection{Smoothness}
	It is needed to consider the complex plane, given by its real and imaginary parts, as a two-dimensional real space, $\mathbb C=\mathbb R\times\mathbb R$.
	Our aim is the proof that $P_{\rm QC}(\boldsymbol\alpha; w)$ is a smooth function, $P_{\rm QC}\in C^\infty(\mathbb C^n,\mathbb R)$.
	In Ref.~\cite{kiesel10}, it was shown for the single-mode case ($n=1$) that the filtered quasiprobability is a square-integrable function.
	Here, we want to go further and prove that the multimode filtered function is smooth for any $w>0$.

	The properties of the characteristic function are basically given by its definition as $\Phi(\boldsymbol\beta)=\langle{:}\hat D(\boldsymbol\beta){:}\rangle$, where
	${:}\hat D(\boldsymbol\beta){:}={\rm e}^{\boldsymbol\beta\cdot\boldsymbol{\hat a^\dagger}}{\rm e}^{-\boldsymbol\beta^\ast\cdot\boldsymbol{\hat a}}$ represents the multimode normally ordered displacement operator (see, e.g.~\cite{VogelWelsch}).
	The relation to the unitary displacement operator is given by ${\rm e}^{-|\boldsymbol\beta|^2/2}\,{:}\hat D(\boldsymbol\beta){:}=\hat D(\boldsymbol\beta)$.
	In addition, we have a continuity in the form $\hat D(\boldsymbol\beta+\boldsymbol\beta')={\rm e}^{-{\rm i}\,{\rm Im}(\boldsymbol\beta'\cdot\boldsymbol\beta^\ast)}\hat D(\boldsymbol\beta)\hat D(\boldsymbol\beta')$.
	Thus, we have that $\Phi(\boldsymbol\beta)$ is continuous and bounded in the form
	\begin{align}
		|\Phi(\boldsymbol\beta)|\leq {\rm e}^{|\boldsymbol\beta|^2/2}.
	\end{align}

	In the following, we need to prove that the Fourier transform of $\Phi_{\rm QC}$ is smooth.
	In addition, we may summarize some properties of the Fourier transform $\mathcal F$.
	The Fourier transform is an automorphism of square-integrable functions $L^2(\mathbb C^{n},\mathbb C)$.
	The inverse Fourier transform of an auto-correlation function is $\mathcal F{\rm Aut}[g]=|\mathcal F g|^2$.
	It is also important to recall the Sobolev's lemma (see, e.g.~\cite{yosida80}):
	If $|\boldsymbol\beta|^{k} g(\boldsymbol\beta)\in L^1(\mathbb C^n,\mathbb C)$ for $k\in\mathbb N$, it follows that the Fourier transform is continuously differentiable, $\mathcal Fg\in C^k(\mathbb C^n,\mathbb C)$.

	\paragraph*{Theorem 1.}
		Let $\omega(\boldsymbol\beta)$ be a positive function, which satisfies $\omega(\boldsymbol\beta){\rm e}^{u |\boldsymbol\beta|^2}\in L^2(\mathbb C^n,\mathbb R)$ for all $u>0$.
		Then $P_{\Omega}(\boldsymbol\alpha; w)\in C^\infty(\mathbb C^n,\mathbb R)$ for all $w>0$.

	\paragraph*{Proof.}
		We have
		\begin{align}
			P_{\rm QC}(\boldsymbol\alpha; w)=\mathcal F[\Phi(\boldsymbol\beta)\Omega(\boldsymbol\beta/w)](\boldsymbol\alpha).
		\end{align}
		It is sufficient to prove that $\Phi_{\Omega}(\boldsymbol\beta;w)=\Phi(\boldsymbol\beta)\Omega(\boldsymbol\beta/w)\in L^2(\mathbb C^n,\mathbb C)$, and $|\boldsymbol\beta|^{2k} \Phi(\boldsymbol\beta)\Omega(\boldsymbol\beta/w)\in L^1(\mathbb C^n,\mathbb C)$ for all $w>0$ and $k\in\mathbb N$.
		We then have
		\begin{align}
			\left||\boldsymbol\beta|^{2k}\Phi_{\rm QC}(\boldsymbol\beta;w)\right|
			\leq& C_u |\boldsymbol\beta|^{2k} {\rm e}^{|\boldsymbol\beta|^2/2}{\rm e}^{-u |\boldsymbol\beta|^2/w^2}\\
			\leq& C_u k!\,{\rm e}^{\left(1+\frac{1}{2}-\frac{u}{w^2}\right)|\boldsymbol\beta|^2}.
		\end{align}
		We can choose a $u>3 w^2/2$, such that the left-hand side is bounded by a rapidly decaying function, which is in $L^1(\mathbb C^n,\mathbb C)\cap L^2(\mathbb C^n,\mathbb C)$.
		\hfill$\square$

	Note that the considered $\omega(\boldsymbol\beta)={\rm e}^{-|\boldsymbol\beta|^4}$ fulfills the requirements.
	We can bound it as $\omega(\boldsymbol\beta){\rm e}^{u |\boldsymbol\beta|^2}={\rm e}^{u^2/4}{\rm e}^{ -(|\boldsymbol\beta|^2-u/2)^2}$.

\section{Sampling error estimation}
	In the following we present the calculation of the empirical variance of the sampling.
	The error estimation for sampling methods of one-mode systems can be found in~\cite{kiesel12a} and the references therein.
	The statistical uncertainties, calculated below, allow us to determine the confidence of any value of the filtered multimode quasiprobability distribution $P_{\rm QC}$.
	
	An estimate for $P_{\rm QC}(\boldsymbol\alpha;w)$ of an $n$-mode field is obtained in terms of a product of the so-called pattern functions,~cf. Eq.~(\ref{Eq:patternfct}), via
	\begin{align}
		P_{\rm QC} (\boldsymbol\alpha;w)=\frac{1}{N} \sum_{j=1}^N \prod_{k=1}^n f(x_k[j],\varphi_k[j],\alpha_k;w).
	\end{align}
	The statistical estimate is based on a sample of a sufficiently large number $N$ of measured quadrature values, $\{( x_k[j],\varphi_k[j])_{k=1}^n\}_{j=1}^N$, for each mode.
	The pair $(x_k[j],\varphi_k[j])$ is the $j$-th measurement outcome for the $k$-th mode, according to the quadrature statistics of the quantum state under study.
	The sampled estimate of $P_{\rm QC} (\boldsymbol\alpha;w)$ is a random variable, because of the large but finite number $N$ of data points per mode. For sufficiently large $N$, it is normally distributed with a small variance.
	The variance of the stochastic quantity $P_{\rm QC}(\boldsymbol\alpha;w)$ is
	\begin{align}
		\sigma^2[P_{\rm QC} (\boldsymbol\alpha;w)]=\mathbb{E}\left[P_{\rm QC} (\boldsymbol\alpha;w)^2 \right]-\mathbb{E}\left[ P_{\rm QC} (\boldsymbol\alpha;w)\right]^2.
	\end{align}
	The expectation value $\mathbb{E}$ is calculated with respect to the normal distribution of $P_{\rm QC}$, considered a sampled stochastic quantity. 

	The first moment is
	\begin{align}
		\mathbb{E}\left[ P_{\rm QC} (\boldsymbol\alpha;w)\right]=& \frac{1}{N} \sum_{j=1}^N \mathbb{E}\left[\prod_{k=1}^n f(x_k[j],\varphi_k[j],\alpha_k;w) \right]=\frac{1}{N} \left\lbrace N\mathbb{E}\left[\prod_{k=1}^n f(x_k,\varphi_k,\alpha_k;w) \right]\right\rbrace.
	\end{align}
	The second moment of $P_{\rm QC}(\boldsymbol\alpha;w)$ reads as
	\begin{gather}
		\mathbb{E}\left[P_{\rm QC} (\boldsymbol\alpha;w)^2 \right]= \mathbb{E}\left[\left(\frac{1}{N} \sum_{j=1}^N \prod_{k=1}^n f(x_k[j],\varphi_k[j],\alpha_k;w) \right) \left(\frac{1}{N} \sum_{j'=1}^N \prod_{k=1}^n f(x_k[j'],\varphi_k[j'],\alpha_k;w) \right)\right]\nonumber\\
		= \frac{1}{N^2} \left\lbrace \sum_{j=1}^N \mathbb{E}\left[\prod_{k=1}^n f^2(x_k[j],\varphi_k[j],\alpha_k;w) \right]+\sum_{\substack{j,j'\\j\neq j'}}^N \mathbb{E}\left[\prod_{k=1}^n f(x_k[j],\varphi_k[j],\alpha_k;w) \right]\mathbb{E}\left[\prod_{k=1}^n f(x_k[j'],\varphi_k[j'],\alpha_k;w) \right]\right\rbrace\nonumber\\
		=\frac{1}{N^2} \left\lbrace N \mathbb{E}\left[\prod_{k=1}^n f^2(x_k,\varphi_k,\alpha_k;w) \right]+N(N-1)\left(\mathbb{E}\left[\prod_{k=1}^n f(x_k,\varphi_k,\alpha_k;w) \right]\right)^2\right\rbrace.
	\end{gather}
	We have decomposed the two original sums into all contributions with $j=j'$ plus $j\neq j'$ and took into account the fact that individual measured points $\{( x_k[j],\varphi_k[j])_{k=1}^n\}_{j=1}^N$ are stochastically independent and equally distributed according to the same Gaussian statistics.
	Finally, the empirical variance is given as 
	\begin{align}
		\sigma^2[P_{\rm QC} (\boldsymbol\alpha;w)]=\frac{1}{N} \left\lbrace \mathbb{E}\left[\prod_{k=1}^n f^2(x_k,\varphi_k,\alpha_k;w)\right]- \left(\mathbb{E}\left[\prod_{k=1}^n f(x_k,\varphi_k,\alpha_k;w)\right]\right)^2\right\rbrace.
	\end{align}
		
	For the particular case of a two-mode system this variance simplifies to
	\begin{align}
		\sigma^2[P_{\rm QC} (\boldsymbol\alpha;w)]=\frac{1}{N} \left\lbrace \mathbb{E}\left[f^2(x_1,\varphi_1,\alpha_1;w)\,f^2(x_2,\varphi_2,\alpha_2;w)\right]-
		\mathbb{E}\left[f(x_1,\varphi_1,\alpha_1;w)\,f(x_2,\varphi_2,\alpha_2;w)\right]^2\right\rbrace.
	\end{align}
	In addition, for our example of a fully phase randomized two-mode squeezed vacuum state, the phase-independent pattern function is of the form
	\begin{align}
		f(x,\alpha;w)=\frac{2}{\pi} \int_0^\infty db\, b\, e^{b^2/2} e^{ibx} J_0(2b|\alpha_1|) \,\Omega\left(b/w\right). 
	\end{align}
	Hence, the first and second moment of the regularized $P$~function are
	\begin{align}\label{Eq:ExpValue}
		\mathbb{E}\left[f(x_1,\alpha_1;w)\,f(x_2,\alpha_2;w)\right]=&\left(\frac{2}{\pi}\right)^2 \int_0^\infty db\, b\, J_0(2b|\alpha_1|) \,\Omega\left(b/w\right)  \int_0^\infty db'  b'\, J_0(2b'|\alpha_2|) \,\Omega\left(b'/w\right) \Phi(b,b'),
		\end{align}
	which is $P_{\rm QC} (\alpha_1,\alpha_2;w)$,
	and
	\begin{align}\label{Eq:SecMoment}
		\mathbb{E}\left[f^2(x_1,\alpha_1;w)\,f^2(x_2,\alpha_2;w)\right]=\left(\frac{2}{\pi}\right)^4 \int_0^\infty db_1 b_1\, J_0(2b_1|\alpha_1|) \,\Omega\left(b_1/w\right)  \int_0^\infty db'_1  b'_1\, J_0(2b'_1|\alpha_1|) \,\Omega\left(b'_1/w\right)  \\
		\int_0^\infty db_2 b_2\, J_0(2b_2|\alpha_2|) \,\Omega\left(b_2/w\right) \int_0^\infty db'_2 b'_2\, J_0(2b'_2|\alpha_2|) \,\Omega\left(b'_2/w\right) e^{b_1b'_1+b_2b'_2} \Phi(b_1-b'_1,b_2-b'_2)\nonumber.
	\end{align}
	
	The characteristic function $\Phi(b,b')$, describing the statistical uncertainties of the measured quadrature values,
 	is explicitly given by
	\begin{align}\label{Eq:cf}
		\Phi(b,b')=e^{b^2/2}e^{b'^2/2}\,\mathbb{E}\left[ e^{ib x_1}e^{ib' x_2} \right]
		=e^{\frac{1}{2}\mathbf{b}^T (I-\mathbf{\Sigma})\mathbf{b} +i\mathbf{b}^T\boldsymbol\mu},
	\end{align}
	where
	\begin{align}
	\mathbf{b}=
		\begin{pmatrix}
			b\\
			b'
		\end{pmatrix}, \qquad 
		\mathbf{x}=
		\begin{pmatrix}
			x_1\\
			x_2
		\end{pmatrix} 
		\qquad \text{and} \qquad 
		\boldsymbol\mu=
		\begin{pmatrix}
			\mu_1\\
			\mu_2
		\end{pmatrix} \qquad \text{with} \qquad  \mu_i=\frac{1}{N} \sum_{j=1}^N x_i[j].
	\end{align}
	The covariance matrix is obtained from the measured data as
	\begin{align}
	\mathbf{\Sigma}=\frac{1}{N}
		\begin{pmatrix}
			\sum_{j=1}^N (x_1[j]-\mu_1)^2 & \sum_{j=1}^N (x_1[j]-\mu_1)(x_2[j]-\mu_2) \\
			\sum_{j=1}^N (x_2[j]-\mu_2)(x_1[j]-\mu_2) & \sum_{j=1}^N (x_2[j]-\mu_2)^2
		\end{pmatrix}.
	\end{align}
	The variance, applied in the calculation of the confidence, is obtained from Eqs. (\ref{Eq:ExpValue}) to (\ref{Eq:cf}).
\end{widetext}


\begin{thebibliography}{99}
	\bibitem{glauber63} R. J. Glauber, Phys. Rev. \textbf{131}, 2766 (1963).
	\bibitem{sudarshan63} E. C. G. Sudarshan, Phys. Rev. Lett. \textbf{10}, 277 (1963).
        \bibitem{titulaer65} U. M. Titulaer and R. J. Glauber, Phys. Rev. \textbf{140}, B676 (1965).
	\bibitem{mandel86} L. Mandel, Phys. Scr. T {\bf T12}, 34 (1986).
	\bibitem{horodecki09} R. Horodecki, P. Horodecki, M. Horodecki and K. Horodecki, Rev. Mod. Phys. {\bf 81}, 865 (2009).
	\bibitem{guehne09} O. G\"uhne and G. T\'oth, Phys. Rep. {\bf 474}, 1 (2009).
	\bibitem{werner89} R. F. Werner, Phys. Rev. A \textbf{40}, 4277 (1989).
	\bibitem{sperling09} J. Sperling and W. Vogel, Phys. Rev. A \textbf{79}, 042337 (2009).
	\bibitem{sperling12} J. Sperling and W. Vogel, New J. Phys. \textbf{14}, 055026 (2012).
	\bibitem{ollivier01} H. Ollivier and W. H. Zurek, Phys. Rev. Lett. {\bf 88}, 017901 (2001).
	\bibitem{henderson01} L. Henderson and V. Vedral, J. Phys. A 34, 6899 (2001).
	\bibitem{ferrarro12} A. Ferraro and M. G. A. Paris, Phys. Rev. Lett. {\bf 108}, 260403 (2012).
	\bibitem{vogel08} W. Vogel, Phys. Rev. Lett. {\bf 100}, 013605 (2008). 
	\bibitem{agarwal92} G.~S.~Agarwal, and K.~Tara, Phys. Rev. A {\bf 46}, 485 (1992).
	\bibitem{kiesel08} T. Kiesel, W. Vogel, V. Parigi, A. Zavatta and M. Bellini, Phys. Rev. A 78, 021804(R) (2008).
	\bibitem{cahill69} K. E. Cahill and R. J. Glauber, Phys. Rev. \textbf{177}, 1882 (1969).
	\bibitem{smithey93} D. T. Smithey, M. Beck, M. G. Raymer, and A. Faridani, Phys. Rev. Lett. {\bf 70}, 1244 (1993).     
	\bibitem{dunn95} T. J. Dunn, I. A. Walmsley, and S. Mukamel, Phys. Rev. Lett. {\bf 74}, 884–887 (1995).
	\bibitem{leibfried96} D. Leibfried, D. M. Meekhof, B. E. King, C. Monroe, W. M. Itano, and D. J. Wineland, Phys. Rev. Lett. {\bf 77}, 4281 (1996).
	\bibitem{ourjoumtsev06} A. Ourjoumtsev, R. Tualle-Brouri, J. Laurat and P. Grangier, Science {\bf 312}, 83 (2006).
	\bibitem{Deleglise08} S. Del\'eglise, I. Dotsenko, C. Sayrin, J. Bernu, M. Brune, J.-M. Raimond, and S. Haroche, Nature {\bf 455}, 510 (2008).
	\bibitem{agarwal70} G. S. Agarwal and E. Wolf, Phys. Rev. D \textbf{2}, 2161, 2187, 2206 (1970).
	\bibitem{kiesel10} T. Kiesel and W. Vogel, Phys. Rev. A \textbf{82}, 032107 (2010).
	\bibitem{kiesel11} T. Kiesel, W. Vogel, M. Bellini, and A. Zavatta, Phys. Rev. A {\bf83}, 032116 (2011).
	\bibitem{kiesel11a} T. Kiesel, W. Vogel, B. Hage and R. Schnabel, Phys. Rev. Lett. {\bf 107}, 113604 (2011).
	\bibitem{kiesel12} T. Kiesel, W. Vogel, S. L. Christensen, J.-B. B{\'e}guin, J. Appel, and E. S. Polzik, Phys. Rev. A {\bf 86}, 042108 (2012).
        \bibitem{datta10} A. Datta, e-print arXiv:1003.5256.
	\bibitem{vogel00} W. Vogel, Phys. Rev. Lett. {\bf 84}, 1849 (2000). 
	\bibitem{Ri-Vo02} Th. Richter, W. Vogel, Phys. Rev. Lett. {\bf 89}, 283601 (2002).
	\bibitem{CF-exp1} A. I. Lvovsky and J. H. Shapiro, Phys. Rev. A {\bf 65}, 033830 (2002). 
	\bibitem{CF-exp2} A. Zavatta,V. Parigi, and M. Bellini,  Phys. Rev. A {\bf 75}, 052106 (2007).
	\bibitem{kiesel09} T. Kiesel, W. Vogel, B. Hage, J. DiGuglielmo, A. Samblowski and R. Schnabel, Phys. Rev. A 79, 022122 (2009).
	\bibitem{giorda} P. Giorda and M. G. A. Paris, Phys. Rev. Lett. {\bf 105}, 020503 (2010).
 	\bibitem{bochner33} S. Bochner, Math. Ann. {\bf 108}, 378 (1933).
	\bibitem{VogelWelsch} W. Vogel and D.-G. Welsch, {\it Quantum Optics}, 3rd ed. (Wiley-VCH, Weinheim, 2006).
	\bibitem{yosida80} K. Yosida, {\it Functional Analysis} (Springer-Verlag, New York, 1980).
	\bibitem{kiesel12a} T. Kiesel, Phys. Rev. A {\bf 85}, 052114 (2012).
\end{thebibliography}
\end{document}